\documentclass[aps,pra,showpacks,twocolumn]{revtex4}

\def\PRA{{Phys.~Rev.~A} }

\def\JPB{{J.~Phys.~B} }
\def\PRL{{Phys.~Rev.~Lett.} }

\newcommand{\myscaleboxa}[1]{\scalebox{0.30}[0.30]{#1}}

\usepackage{graphicx}
\begin{document}


\title{Retrieval of electron-atom scattering cross sections from laser-induced electron rescattering of atomic negative ions in intense laser fields}

\author{XiaoXin Zhou$^{1,2}$, Zhangjin Chen$^{1}$, Toru Morishita$^{1,3}$, Anh-Thu
Le$^{1}$, and C. D. Lin$^{1}$ }

\affiliation{ $^1$J. R. Macdonald Laboratory, Physics Department,
Kansas State University, Manhattan, Kansas 66506-2604 USA\\
$^2$College of Physics and Electronic Engineering, Northwest Normal
University, Lanzhou 730070, P. R. China\\
$^3$Department of Applied Physics and Chemistry, University of
Electro-Communications, Tokyo, 182-8585, Japan and PRESTO, Japan
Science and Technology Agency, Kawaguchi, Saitama, 332-0012, Japan}

\date{\today}

\begin{abstract}

We investigated the two-dimensional electron momentum distributions
of atomic negative ions in an intense laser field by solving the
time-dependent Schr?dinger equation (TDSE) and using the first- and
2nd-order strong-field approximations (SFA). We showed that
photoelectron energy distributions and low-energy photoelectron
momentum spectra predicted from SFA are in reasonable agreement with
the solutions from the TDSE.  More importantly, we showed that
accurate electron-atom elastic scattering cross sections can be
retrieved directly from high-energy electron momentum spectra of
atomic negative ions in the laser field. This opens up the
possibility of measuring electron-atom and electron-molecule
scattering cross sections from the photodetachment of atomic and
molecular negative ions by intense short lasers, respectively, with
temporal resolutions in the order of femtoseconds.

\end{abstract}


\maketitle

\section{Introduction}

In the past two decades, the energy and momentum spectra of
electrons generated by the ionization of atoms or detachment of
negative ions by intense laser pulses have been widely investigated
\cite{milosevic_JPB06, Agostini}. In particular, the detachment of
negative hydrogen and fluorine ions in intense laser fields have
been reported in experiments \cite{Reichle_PRL01, Reichle_PRA03,
Kiyan_PRL03, Bergues_PRL05, Bergues_PRA07} and in theory
\cite{milosevic_PRA03, Busuladzic_PRA04, Bivona_PRA07, Reiss_PRA07,
Starace_PRL03, Starace_PRA06, Arendt_PRA07}, using either the
strong-field approximation (SFA) or by solving the time-dependent
Schr?dinger equation (TDSE). These studies have shown that
experimentally observed spectra are largely well reproduced by
theoretical calculations. However, the electrons measured in these
experiments tend to be restricted to low energies, and SFA and TDSE
calculations are often carried out by separate groups using
different additional approximations.

In this paper, we studied the photodetachment of H$^-$ and F$^-$
negative ions in short intense laser pulses. Our goal is to examine
the electron energy distributions and two-dimensional (2D) electron
momentum spectra over a broad energy range, from the threshold up to
10 UP, where UP is the ponderomotive energy.  We will use the
standard first-order SFA (SFA1) and the second-order SFA (SFA2) to
describe the photodetachment process, as well as obtaining the same
spectra by solving the TDSE directly. The negative ion will be
approximated by a model potential and the same potential is used in
the SFA and TDSE calculations. We will establish, based on specific
numerical results, the accuracy of the SFA for describing the
photodetachment of negative ions by lasers. In SFA, it was assumed
that electrons released into the continuum can be described as a
free electron in the laser field, and that there are no excited
states in the target atom.  Both approximations are expected to work
better for negative ion targets than for atomic targets. Indeed our
calculations showed that the total electron energy spectra and the
electron momentum spectra at low energies calculated from the SFA
are in quite good overall agreement with the TDSE results, to within
about a factor of two. Such agreement has not been seen if the
targets are neutral atoms or positive ions \cite{Chirila}. Our major
goal, however, is to establish that  high-energy electron momentum
spectra of negative ions induced by intense short laser pulses can
be used to extract the elastic differential cross sections of
neutral atoms by free electrons. Similar conclusions have been shown
recently for atoms where electron-ion scattering cross sections can
be extracted from laser-induced high-energy electron momentum
spectra of neutral atoms \cite{Toru_PRL07}.

In Section II, we summarize the theoretical models used. The
numerical results for H$^-$ and F$^-$ from SFA and TDSE are
presented and analyzed in Section III, for both the energy
distributions and the 2D momentum distributions. From the
high-energy 2D photoelectron momentum distributions we will extract
the e-H and e-F elastic scattering cross sections.  In Section IV we
summarize the results and discuss the important possible
applications of using laser-induced photoelectron spectra for
probing time-resolved electron-atom or electron-molecule collisions.
Atomic units are used throughout, unless otherwise indicated.

\section{Theoretical Methods}

We will model each negative ion in the single-active electron (SAE)
approximation. In this model, the time-dependent wavefunction of the
active electron in the laser field is governed by
\begin{eqnarray}
i\frac{\partial}{\partial
t}\psi(\textbf{r},t)=[H_0+H'(t)]\psi(\textbf{r},t)
\end{eqnarray}
where $H_0=-\frac{1}{2}\nabla^2+V(\textbf{r})$ is the electron
Hamiltonian of the laser-free system, with $V(\textbf{r})$ being the
atomic model potential, and $H'(t)=\textbf{r}\cdot\textbf{E}(t)$ is
the interaction of the electron with the laser's electric field
$\textbf{E}(t)$. Equation (1) can be solved numerically by using the
split-operator method \cite{chen06, Toru_PRA07}. The electron
momentum spectra are obtained by projecting out the final
time-dependent wavefunction after the laser pulse is over using the
continuum scattering eigenstates of the laser-free Hamiltonian. For
electrons with momentum $p$ in the direction $\hat{\textbf{p}}$,
with respect to the laser polarization direction, if the detachment
amplitude is $f$, then the angular distributions at energy $E=p^2/2$
is given by
\begin{eqnarray}
\frac{\partial P}{\partial {\textbf{p}}}=|f(\textbf{p})|^2.
\end{eqnarray}
This equation can also be rewritten in terms of 2D momentum
distributions. By integrating over the directions of photoelectrons
at a fixed energy $E$, the energy dependence of the photoelectrons
are calculated from
\begin{eqnarray}
\label{energy}\frac{\partial P}{\partial E}=\int |f(\textbf{p})|^2 p
d\hat{\textbf{p}}.
\end{eqnarray}
In the strong-field approximation, the detachment probability of
electrons with momentum ${\textbf{p}}$ is expressed as \cite{chen07}
\begin{eqnarray}
\label{full_SFA}f(\textbf{p}) =f^{(1)}+f^{(2)}
\end{eqnarray}
where the first-order of SFA amplitude (SFA1) is
\begin{eqnarray}
\label{1st-order}f^{(1)}=-i
\int_{-\infty}^{\infty}dt\left\langle\chi_{\textbf{p}}(t)\left|H'(t)\right|\Psi_{0}(t)\right\rangle
\end{eqnarray}
and the second-order of SFA amplitude (SFA2) is
\begin{eqnarray}
\label{2nd-order}f^{(2)}&=&-\int_{-\infty}^{\infty}dt\int_{-\infty}^{t}dt^{\prime}\int
d\textbf{k}\left\langle\chi_{\textbf{p}}(t)\left|V\right|\chi_{\textbf{k}}(t)\right\rangle
\nonumber \\ && \times
\left\langle\chi_{\textbf{k}}(t^{\prime})\left|H'(t^{\prime})\right|\Psi_{0}(t^{\prime})\right\rangle
\end{eqnarray}
Here $\chi_{\textbf{p}}(t)$ and $\Psi_{0}(t^{\prime})$ are the
Volkov states of a continuum electron with momentum $\textbf{p}$ and
the initial state, respectively. The evaluation of the integral
(\ref{2nd-order}) has been discussed in \cite{chen07}. In SFA2, we
have used the saddle point approximation in evaluating the
integration over $d\textbf{k}$ where $\textbf{k}$ is the momentum of
photoelectron in the intermediate Volkov state. Otherwise all the
integrals are evaluated numerically without additional
approximations.

For the model potential, for H$^-$, we take it from \cite{Chu}:
\begin{eqnarray}
\label{pot_H_min}V(r)=-(1+\frac{1}{r})e^{-2r}-\frac{\alpha_d}{2r^4}(1-e^{(r/r_c)^6})+V_0(r)
\end{eqnarray}
where $V_0(r)=(c_0+c_1r+c_2r^2)e^{-\beta r}$.

Note that the polarization of the core (H atom) has been included in
(\ref{pot_H_min}), with $\alpha_d$ and $r_c$ being the
polarizability and the cutoff constant, respectively. The parameters
in (\ref{pot_H_min}) are listed in Ref. \cite{Chu}. For the negative
fluorine ion, we chose the potential to have the form,
\begin{eqnarray}
\label{pot_F_min}V(r)=-a_1\frac{e^{-\alpha_1
r}}{r}-a_2\frac{e^{-\alpha_2 r}}{r}.
\end{eqnarray}
From fitting, the parameters $a_1$, $a_2$, $\alpha_1$ and $\alpha_2$
obtained are 5.137, 3.863, 1.288 and 3.545 respectively. This
potential gives the correct ground state energy of the negative
fluorine ion, and its lowest p-orbital eigenfunction agrees well
with the tabulated Hartree-Fock wavefunction for F$^-$.

\section{Results and Discussion}

We used the TDSE and SFA to calculate the energy spectra and the 2D
momentum distributions of H$^-$ and F$^-$ in intense laser fields.
For H$^-$, we took peak laser intensity of $I=1.0\times10^{11}$
W/cm$^2$, wavelength $\lambda=10600$ nm and pulse duration of 3
cycles. The electric field has the form $E(t)=E_0F(t)\cos(\omega
t+\phi)$ where $\phi$ (we chose it to be zero) is the carrier
envelope phase, and $F(t)=\cos^2(\pi t/\tau)$ where $\tau=3T$
($|t|\leq \tau/2$), with T being the period of the laser. For this
laser pulse, the ponderomotive energy is 1.05 eV, and it needs 16
photons ($\hbar\omega=0.117$ eV) to remove the electron from H$^-$.
The Keldysh parameter is 0.6, thus the electron detachment is well
in the tunnelling regime.

\begin{figure}
\mbox{\rotatebox{270}{\myscaleboxa{
\includegraphics{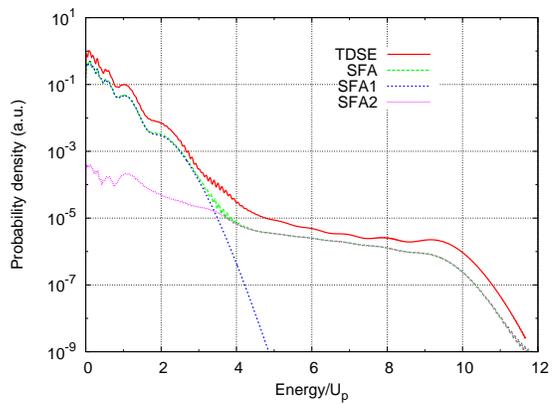}}}}
\caption{(Color online) Energy spectra of laser-induced
photoelectrons of H$^-$ obtained from TDSE (red solid line) and from
SFA (green dotted line). The laser parameters are
$I=1.0\times10^{11}$ W/cm$^2$ and $\lambda=10600$ nm for a pulse
with 3 optical cycles (see text).}
\end{figure}

In Fig. 1 we show the calculated energy spectra of the detached
photoelectrons where the energy is expressed in units of $U_p$. In
the SFA, the low-energy part below about 3.5 $U_p$ is dominated by
SFA1, while the yield above $4U_p$ is dominated almost entirely by
SFA2, i.e., it is due to the recollision between the detached
electrons with the neutral target H.  The figure also shows that the
energy spectra obtained from SFA are only about a factor of two
smaller than those obtained from the TDSE. For neutral atom or
positive ion targets the absolute total ionization yields calculated
using SFA are generally much smaller, close to one order or more
smaller than those obtained from TDSE \cite{Chirila}.

\begin{figure}
\mbox{\rotatebox{270}{\myscaleboxa{
\includegraphics{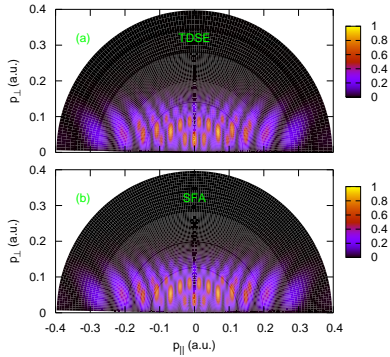}}}}
\caption{(Color online) 2D momentum distributions of laser-induced
photoelectrons of H$^-$ in the low energy region, obtained from TDSE
(a) and from SFA (b). The horizontal axis is the electron momentum
along the direction of laser polarization and the vertical axis is
any axis perpendicular to it (the electron spectrum has cylindrical
symmetry.) The laser parameters are the same as in Fig. 1. }
\end{figure}

Fig. 2 shows the 2D momentum distributions of the detached electrons
at low energies where $p_{||}$  refers to the direction of the laser
polarization and $p_{\bot}=\sqrt{p^2-p_{||}^2}$ is perpendicular to
it. We normalized the data to the same peak value in the two graphs.
There are several special features that should be mentioned. First,
the 2D spectra predicted from SFA1 and from TDSE are very close to
each other. Second, the 2D spectra exhibit clean structures that may
be attributed to multiphoton detachment (MPD), despite that it is in
the tunneling regime. It is interesting to compare these low-energy
2D momentum spectra with those from neutral atomic targets
\cite{chen06}. To begin with, for atomic targets, the low-energy
momentum spectra show fan-like structure emanating from the origin.
These fan-like structures have been observed experimentally
\cite{Cocke_JPB06, Ullrich_JPB04}, and are seen only from the TDSE
calculations, not from the SFA1 calculations. The fan-like
structures had been attributed to the long-range Coulomb interaction
between the electron and the positive atomic ion after the atom is
ionized by the laser \cite{chen06, Arbo_PRL06}. For negative ion
targets, there is no long-range Coulomb interaction between the
electron and the neutral atom following the detachment of the
electron by the laser, thus the fan-like structure is absent in the
TDSE results, as shown in Fig. 2(a). Figs. 2(a) and (b) both show
clear MPD peaks. The angular distribution of the lowest MPD peak has
a maximum at 90$^\circ$ (even parity), thus the dominant angular
momentum quantum number should be even. For atomic targets, the
dominant angular momentum of the photoelectrons at low energies can
be calculated from the propensity rule given in Chen \emph{et al}.
\cite{chen06}. Since the minimum number of photons needed to remove
the electron at the laser intensity used is 16 and that the ground
state of H? is an s-orbital, Fig. 8 of Chen \emph{et al}.
\cite{chen06} shows that the dominant angular momentum is six, and
thus the parity is even, in agreement with the calculated results
shown in Fig. 2. Note that the next MPD peak would have odd parity
since one more photon is absorbed, as clearly seen from the
calculated spectra.

\begin{figure}
\mbox{\rotatebox{270}{\myscaleboxa{
\includegraphics{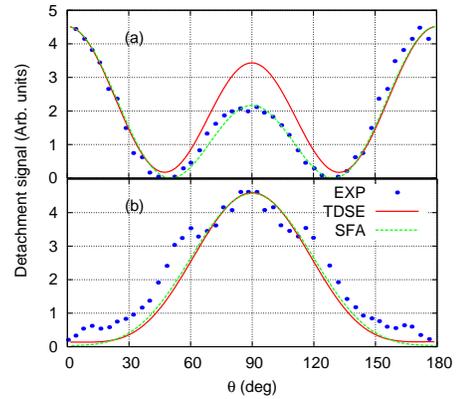}}}}
\caption{(Color online) Angular distributions of laser-induced
photoelectrons of H$^-$, generated by a laser with (a) wavelengh
$\lambda=2150$ nm, intensity $I=1.3\times10^{11}$ W/cm$^2$, for
electron energy $E=0.34$ eV; (b) for intensity $I=6.5\times10^{11}$
W/cm$^2$, and electron energy $E=0.12$ eV. The TDSE and SFA results
are compared to the experimental data of Ref. \cite{Reichle_PRL01}.}
\end{figure}

Low-energy photoelectron angular distributions of H$^-$ have been
reported by Reichle \emph{et al}. \cite{Reichle_PRL01} using lasers
with  $\lambda=2150$ nm ($\hbar\omega=0.576$ eV), pulse duration
(FWHM ) of  250 fs, at two peak intensities $I=1.3\times10^{11}$
W/cm$^2$ and $I=6.5\times10^{11}$ W/cm$^2$, corresponding to Keldysh
parameters of 2.6 and 1.2, respectively. Unlike the previous
example, this is in the multiphoton regime. In fact, the first
low-energy photoelectron peak can be attributed to two-photon
absorption. In Fig. 3 we show the angular distributions for
photoelectron energy at 0.34eV for $I=1.3\times10^{11}$ W/cm$^2$ and
at 0.12 eV for $I=6.5\times10^{11}$ W/cm$^2$. We compare the
experimental data with SFA1 and with TDSE calculations. Both the
SFA1 and the TDSE results agree well with the experimental results
[the actual values from SFA1 shown have been scaled to agree with
the TDSE results at small angles in (a) and at 90$^\circ$ in (b)].
Note that it appears that the SFA1 results are in better agreement
with the experimental data in Fig. 3(a) than the TDSE results.
However, the volume effect is not included in our calculations and
we considered the agreement satisfactory.  Our calculations also
showed that the angular distributions are independent of the pulse
durations.

\begin{figure}
\mbox{\rotatebox{270}{\myscaleboxa{
\includegraphics{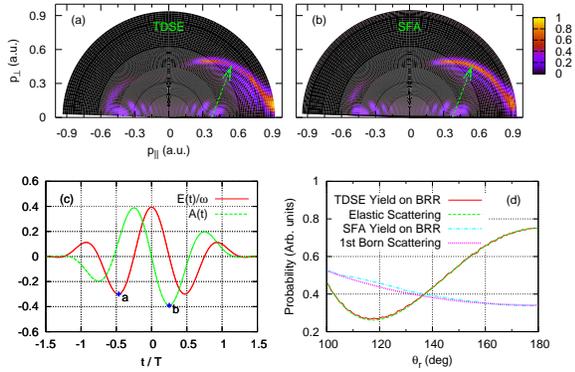}}}}
\caption{(Color online) (a,b): 2D momentum distributions of
laser-induced photoelectrons for H$^-$ in the high energy region,
obtained from TDSE and SFA, respectively; (c) The electric field and
the vector potential of the 3-cycle pulse are depicted, indicating
the "born time" of the detached electron and its return time;(d)
Elastic scattering cross sections extracted along the BRR from the
TDSE calculations (red solid line), and are compared to the
electron-H scattering cross sections (green dotted lines). Also
shown are the corresponding results from SFA and from FBA.}
\end{figure}

Next we return to the laser parameters used in Figs. 1 and 2, and
study the 2D momentum distributions of the detached electrons in the
high energy region. It is well known that these high energy
electrons are generated by the rescattering processes. In order to
highlight these high-energy features, in Figs. 4(a) and 4(b) we show
the 2D momentum distributions by renormalizing them such that the
total ionization probability density at each energy is the same. The
results from the TDSE and from SFA appear to be quite similar
globally. Here we focus on the outermost ring of the 2D spectra. The
distributions along this ring have been studied for laser-generated
photoelectrons from neutral atoms recently \cite{Toru_PRL07}. They
are called back rescattered ridge (BRR) electrons, representing
returning electrons that have been rescattered into the backward
directions by the target ion. The BRR electrons in Fig. 4(a) or 4(b)
are electrons that have been rescattered into the backward
directions by the neutral atomic hydrogen. In Fig. 4(c), we depict
the electric field and the vector potential  of a three-cycle laser
pulse used in the calculation. Electrons that are released at the
half-cycle centered at "a" [see Fig. 4(c)] will return near "b",
with peak kinetic energy at $p_r^2/2=3.17\bar{U}_p$ where
$\bar{U}_p=A_r^2/4$, with $A_r$ being the vector potential at "b".
An electron that is scattered into the backward direction with
momentum $p_r\hat{\textbf{p}}_r$ at "b" will gain additional
momentum $A_r$ as it emerges from the laser field. Thus the momentum
of the BRR electrons at the end of the laser pulse is given by
$\label{BRR}{\textbf{p}}=-{\textbf{A}}_r+{\textbf{p}}_r$, or
$\label{BRR1}p_{||} =-A_r-p_r\cos\theta_r$ and $
p_{\perp}=p_r\sin\theta_r$, where $p_r=1.26A_r$. Here $\theta_r$ is
the scattering angle measured from the "incident" direction of the
recolliding electrons and for BRR, we consider $\theta_r$ larger
than only. (The BRR ring is indicated by showing an arrow with
length from its shifted center in Figs. 4(a) and 4(b).) In Fig. 4(d)
we show the yield of the photoelectrons along the BRR, plotted
against the scattering angle $\theta_r$. On the same graph we also
show the differential elastic cross sections of electrons scattered
by atomic hydrogen. For the laser parameters used in this example,
$A_r=0.39$ and thus the electron's kinetic energy is  $p_r^2=3.28$
eV. In Fig. 4(d), the yields along the BRR obtained from the TDSE
calculations are compared to "exact" electron-hydrogen differential
scattering cross sections (normalized), and they are in good
agreement. On the same graph, the yields from the SFA2 calculations
are compared to the differential elastic e-H scattering cross
sections calculated in the first Born approximation (FBA) and they
also show good agreement between them. In the SFA2, the returning
electron interacts with the H atom only once and thus the elastic
scattering cross section is given by the FBA.  The yields along the
BRR obtained from the TDSE solution and from SFA2 are clearly
different. In particular, the extracted elastic differential cross
section from TDSE shows a minimum and becomes larger at large
angles. The one extracted from SFA2 or from FBA show that the
differential cross sections decrease monotonically with increasing
angles. We thus confirm that one can extract differential elastic
e-H scattering cross sections in the backward directions from
laser-induced photodetachment of H? ions by intense short laser
pulses.

\begin{figure}
\mbox{\rotatebox{270}{\myscaleboxa{
\includegraphics{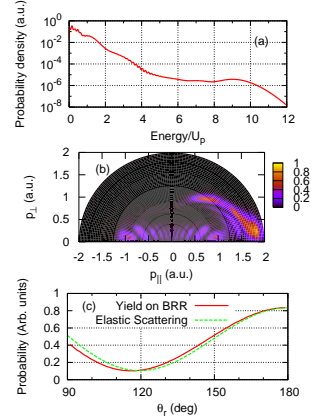}}}}
\caption{(Color online) (a) Energy spectra of laser-induced
photoelectrons of F$^-$ obtained from solving the TDSE. The laser
parameters are $I=1.3\times10^{13}$ W/cm$^2$ and $\lambda=1800$ nm
for a pulse with 3 optical cycles; (b) 2D electron momentum
distributions in the high energy region; (c) Elastic scattering
cross sections, extracted along the BRR from the TDSE (red solid
line) calculation, as compared to the e-F elastic scattering cross
sections (green dotted line).}
\end{figure}

We have carried out similar studies for F$^-$ ions, using a 3-cycle
laser pulse with wavelength $\lambda=1800$ nm and peak intensity
$I=1.3\times10^{13}$ W/cm$^2$. The ponderomotive energy $U_p$ is
3.93 eV. In this case, it needs 11 photons to remove the electron
from F$^-$. We show the energy spectra of the detached electron from
F$^-$ ions in Fig. 5(a).  In Fig. 5(b), we show the 2D momentum
distributions of the detached electrons for F$^-$ calculated using
TDSE, and focusing on electrons along the BRR. The yields along the
BRR show a clear minimum. This minimum is due to the minimum in the
differential scattering cross sections between the electrons and
neutral fluorine atoms. For the laser parameters used, we calculated
that the returning electron energy is 11.8 eV. In Fig. 5(c) we show
extracted differential scattering cross sections from laser-induced
momentum distributions along the BRR, and compare the results with
the calculated elastic e-F scattering cross sections. The good
agreement between the two demonstrates once again that one can
extract electron-atom scattering cross sections from laser-induced
photoelectron momentum distributions of negative ions.

\section{Summary and Perspective}

In this paper we studied the photodetachment cross sections of
atomic negative ions by an intense laser pulse, using the
first-order and second-order strong-field approximations (SFA1+
SFA2), and by solving the time-dependent Schr?dinger equation (TDSE)
directly. We established that the electron energy distributions and
the 2D electron momentum distributions obtained from SFA and from
TDSE are quite close to each other. On the other hand, there are
important differences in the 2D electron momentum distributions at
high energies. These electrons are identified to be distributed
along the so-called back rescattered ridge (BRR). From the 2D
momentum distributions along the BRR, we showed that elastic
scattering cross sections between free electrons with neutral atom
targets can be accurately extracted. In other words, we have shown
that specific structure information which has previously been probed
with electron scattering can also be obtained from the
photodetachment of the negative ions by short intense laser pulses.

The results from this work have two important implications. Since
the laser pulses have short durations down to a few femtoseconds,
the present results imply that one can probe the time-evolution of a
transient negative ion system using this method, similar to what one
can do with neutral targets \cite{Toru_PRL07, Toru_NJP07}. Second,
it shows that one can extract electron-atom or electron-molecule
scattering cross sections by carrying out photodetachment
measurements of their negative ions with intense laser pulses. One
can control the electron scattering energies by changing laser's
intensity or wavelength. For many atomic and molecular species,
including radicals, it may be easier to produce or control their
negative ions than their neutrals. In this case photodetachment of
negative ions by intense short laser pulses can potentially offer a
much better means for measuring electron scattering cross sections.

Although we have carried out calculations using negative atomic ions
treated by a model potential, we believe that our conclusions
applies to real many-electron atomic and molecular systems. In fact,
it is well-known that electron scattering off an atom or molecule
cannot be accurately described using a simple model potential at low
incident energies. At present, it is not possible to obtain accurate
2D momentum spectra for many-electron atoms or molecules from TDSE.
If accurate laser-generated electron spectra are available, one
would expect the data to reveal many-electron effects. For example,
in electron-hydrogen atom scattering, in Fig. 4d we show that the
differential cross section has a minimum at 120$^\circ$ according to
the model potential we have used. Experimental data by Williams
\cite{Williams} at $E=3.4$ eV indicated that the minimum occurs
closer to about 80$^\circ$ (At 8.7 eV the minimum is at about
115$^\circ$). To achieve agreement for electron scattering at such
low energies, more advanced theoretical models which treats both
electrons together, such as the Kohn variational principle
\cite{Schwartz}, the polarized orbital method \cite{Temkin} or
close-coupling methods \cite{Burke} are needed. It remains to be
seen if laser-induced photoelectron spectra can measured with high
precision such that many-electron effects are revealed. Unlike
electron scattering which is a linear phenomenon, laser-induced
photoelectron yields are nonlinear phenomena. On the other hand, the
present paper shows that the extracted electron scattering cross
sections from the BRR are independent of the laser parameters,
including their intensity or wavelength (and the duration
\cite{chen07}), so long that the returning electron has the same
energy. This may provide an added check on the electron scattering
cross sections extracted from laser-induced photoelectron
measurements.

\begin{acknowledgements}

This work was supported in part by Chemical Sciences, Geosciences
and Biosciences Division, Office of Basic Energy Science, US
Department of Energy. X. X. Zhou was supported in part by the
National Natural Science Foundation of China (Grant No. 10674112).
TM was supported in part by the PRESTO  program of the Japan Science
and Technology Agency (JST), by a Grants-in-Aid for Scientific
Research from a Japanese Society for the Promotion of Science
(JSPS), by the 21st Century COE program on "Coherent Optical
Science", and by JSPS Bilateral joint program between US and Japan.

\end{acknowledgements}


\begin{thebibliography}{xx}
\bibitem{milosevic_JPB06} D. B. Milo\v{s}evi\'{c}, G. G. Paulus,
D. Bauer and W. Becker, \JPB {\bf 39}, R203 (2006).

\bibitem{Agostini} P. Agostini and L. F. DiMauro, Rep. Prog. Phys. {\bf 67}, 813 (2004).

\bibitem{Reichle_PRL01}R. Reichle, H. Helm and I. Y. Kiyan, \PRL {\bf 87}, 243001
(2001).

\bibitem{Reichle_PRA03} R. Reichle, H. Helm, and I. Y. Kiyan, \PRA {\bf 68}, 063404 (2003).

\bibitem{Kiyan_PRL03} I. Y. Kiyan and H. Helm, \PRL {\bf 90}, 183001 (2003).

\bibitem{Bergues_PRL05} B. Bergues, Y. Ni, H. Helm, and I. Y. Kiyan, \PRL {\bf
95}, 263002 (2005).

\bibitem{Bergues_PRA07} B. Bergues, Z. Ansari, D. Hanstorp, and I. Y. Kiyan, \PRA {\bf
75}, 063415 (2007).

\bibitem{milosevic_PRA03} D. B. Milo\v{s}evi\'{c}, A. Gazibegovic-Busuladzic, and W. Becker,
\PRA {\bf 68}, 050702 (2003).

\bibitem{Busuladzic_PRA04} A. Gazibegovic-Busuladzic, D. B. Milo\v{s}evi\'{c}, and W.
Becker,PRA {\bf 70}, 053403 (2004).

\bibitem{Bivona_PRA07} S. Bivona, G. Bonanno, R. Burlon, and C. Leone, \PRA {\bf 76},
021401(R) (2007).

\bibitem{Reiss_PRA07} H. R. Reiss, \PRA {\bf 76}, 033404 (2007).

\bibitem{Starace_PRL03} M. V. Frolov, N. L. Manakov, E. A. Pronin, and A.
F. Starace, \PRL {\bf 91}, 053003 (2003).

\bibitem{Starace_PRA06} K. Krajewska, II. I. Fabrikant, and A. F. Starace, \PRA {\bf 74},
053407 (2006).

\bibitem{Arendt_PRA07} C. Arendt, D. Dimitrovski, and J. Briggs, \PRA {\bf 76},
023423(2007).

\bibitem{Chirila} C. C. Chirila and R. M. Potvliege, \PRA {\bf 71},
021402 (2005).

\bibitem{Toru_PRL07} T. Morishita, A. T. Le, Z. Chen and C. D. Lin, \PRL in press (2007).

\bibitem{chen06} Z. Chen, T. Morishita, A. T. Le, M. Wickenhauser, X. M. Tong, and C. D. Lin,
\PRA {\bf 74}, 053405 (2006).

\bibitem{Toru_PRA07} T. Morishita, Z. Chen, S. Watanabe, and C. D. Lin,
\PRA {\bf 75}, 023407(2007).

\bibitem{chen07} Z. Chen, T. Morishita, A. T. Le, and C. D. Lin, \PRA {\bf 76},
043402 (2007).

\bibitem{Chu} C. Laughlin and S. I. Chu, \PRA {\bf 48}, 4654(1993).

\bibitem{Cocke_JPB06} C. M. Maharjan, A. S. Alnaser, I. Litvinyuk, P. Ranitovic,
and C. L. Cocke,  \JPB {\bf 39}, 1955 (2006).

\bibitem{Ullrich_JPB04} A. Rudenko, K. Zrost, C. D. Schr?ter, V. L. B. de Jesus,
B. Feuerstein, R. Moshammer, and J. Ullrich, \JPB {\bf 37}, L407
(2004).

\bibitem{Arbo_PRL06} D. G. Arb\'{o}, S. Yoshida, E. Persson, K. I. Dimitriou, and J.
Burgd\"{o}rfer, \PRL {\bf 96}, 143003 (2006).

\bibitem{Toru_NJP07} T. Morishita, A. T. Le, Z. Chen, and C. D. Lin, New J. Phys. in
press (2007).

\bibitem{Williams} J. F. Williams, \JPB {\bf 8}, 1683 (1975).

\bibitem{Schwartz} C. Schwartz, Phys. Rev. {\bf 121}, 788 (1961).

\bibitem{Temkin} A. Temkin, Phys. Rev. {\bf 126}, 130 (1962).

\bibitem{Burke} P. G. Burke, D. F. Gallaher and S. Geltman, \JPB {\bf 2}, 1142
(1969).

\end{thebibliography}
\end{document}